# Human Digital Twins in Personalized Healthcare: An Overview and Future Perspectives

MELVIN MOKHTARI, McMaster University, Canada

## ABSTRACT

Digital twins (DTs) are redefining healthcare by paving the way for more personalized, proactive, and intelligent medical interventions. As the shift toward personalized care intensifies, there is a growing need for an individual's virtual replica that delivers the right treatment at the optimal time and in the most effective manner. The emerging concept of a Human Digital Twin (HDT) holds the potential to revolutionize the traditional healthcare system much like digital twins have transformed manufacturing and aviation. An HDT mirrors the physical entity of a human body through a dynamic virtual model that continuously reflects changes in molecular, physiological, emotional, and lifestyle factors. This digital representation not only supports remote monitoring, diagnosis, and prescription but also facilitates surgery, rehabilitation, and overall personalized care, thereby relieving pressure on conventional healthcare frameworks. Despite its promising advantages, there are considerable research challenges to overcome as HDT technology evolves. In this study, I will initially delineate the distinctions between traditional digital twins and HDTs, followed by an exploration of the networking architecture integral to their operation—from data acquisition and communication to computation, management, and decision-making—thereby offering insights into how these innovations may reshape the modern healthcare industry.

## 1  INTRODUCTION

Digital twins (DTs) represent a transformative technology that has evolved significantly, emerging as digital replicas of physical entities, including machinery, devices, and human beings. This evolution indicates an expansion from industrial uses into diverse fields, including healthcare [61], [59]. The core functionalities of digital twins include an accurate mirroring of their physical counterparts, capturing all associated processes in a data-driven manner, maintaining a continuous connection that synchronizes with the real-time state of their physical twins, and simulating physical behavior for predictive analysis [85].

In the context of healthcare, a novel extension of this technology manifests in the form of Human Digital Twins (HDTs), designed to provide a comprehensive digital mirror of individual patients. HDTs not only represent physical attributes but also integrate dynamic changes across molecular, physiological, and behavioral dimensions. This advancement is aligned with a shift toward personalized healthcare (PH) paradigms, enabling tailored treatment strategies based on a patient's unique health profile, thereby enhancing preventive, diagnostic, and therapeutic processes in clinical settings [44], [50]. The personalization aspect of HDTs underscores their potential to revolutionize healthcare by facilitating precise and individualized treatment plans that optimize patient outcomes [72].

Although the potential of digital twins in healthcare has garnered much attention, practical applications remain newly developing, with critical literature highlighting that many implementations are still in exploratory stages [59]. Notably, institutions like the IEEE Computer Society and Gartner recognize this technology as a pivotal component in the ongoing evolution of healthcare systems that emphasize both precision and personalization [31], [89]. The capacity

---

Author's address: Melvin Mokhtari, McMaster University, Hamilton, Canada, melvim1@mcmaster.ca.

of digital twins extends beyond simulation; they provide critical analytics to support clinical decision-making and operational management, representing a paradigm shift towards data-informed healthcare practices [37].

The pursuit of developing personalized digital models within healthcare is driven by several complexities. Firstly, the inherent individuality among patients complicates the calibration of digital models due to diverse health variations, necessitating an intricate understanding and incorporation of personal data into treatment modalities that standard machine models cannot achieve [40]. Secondly, the absence of a universal framework poses significant obstacles, as healthcare systems often lack cohesive methodologies that integrate diagnostics, therapy selection, treatment planning, and patient wellness into a singularly cohesive operational structure [17], [84].

Furthermore, the lack of established reference standards in PH creates a gap in approaches that can efficiently encompass diverse medical areas, thus limiting the development and implementation of effective HDTs. Most organizational efforts remain fragmented, focusing solely on isolated aspects of personalized medicine or chronic disease management, and often neglecting the holistic integration required for effective DT frameworks [65]. The integration of advanced analytics and machine learning (ML) technologies is critical for overcoming these limitations, providing a robust backbone for the data-driven requirements essential for personalized healthcare solutions [43].

This research will explore core technologies necessary for the effective deployment of HDTs in PH applications. It will differentiate HDTs from their traditional counterparts, exemplifying current applications of digital twins within various healthcare spheres, including precision medicine, clinical trial design, and operational efficiencies in hospital settings [23], [46]. I provide an overview of the HDT's networking architecture, which includes layers for data acquisition, data communication, computation, data management, and data analysis and decision-making. Through this examination, existing opportunities alongside barriers to the broader application of HDTs will be delineated.

Ultimately, the objective is to provide a foundational roadmap that elucidates the path toward navigating the complexities inherent in the adoption of HDTs, subsequently unlocking their full potential as transformative innovations in health systems [83].

## 2   HUMAN DIGITAL TWIN FOR PERSONALIZED HEALTHCARE

Digital health innovations are significantly reshaping our healthcare systems by integrating a variety of advanced technologies that foster the delivery of more accurate, personalized, and proactive care [12]. At the forefront of these innovations, digital sensors and IoT devices play a crucial role by continuously capturing data from patients and their environments. This data is transmitted to the cloud, where sophisticated big data tools and analytics extract actionable insights. Such a near-real-time flow of information is essential for creating a dynamic digital replica of a patient, effectively reflecting various stages of disease progression and providing clinicians with a deeper understanding of a patient's medical journey [110].

Artificial intelligence (AI) is increasingly vital in this healthcare transformation [12], [13]. AI systems process the massive amounts of data generated by digital health tools and, crucially, learn from this data to enhance individualized care. Through complex algorithms, AI can reproduce human cognitive processes to address intricate diagnostics and formulate personalized treatment plans. However, a significant drawback in deploying AI within healthcare is the necessity for large, high-quality datasets for training these systems, which can be cumbersome and costly to compile due to the extensive efforts required to manually label individual medical records. Consequently, many AI-driven healthcare solutions are constrained by traditional regression and classification methods that tend to focus on specific diseases or population subsets [13].

To address these challenges, digital twin technology has been introduced as a promising solution [28]. A digital twin—a virtual model reflecting an individual's health—facilitates the simulation of countless "what if" scenarios safely, allowing researchers and clinicians to explore potential outcomes without risk to the patient. This technology significantly enhances AI capabilities by providing diverse, synthetic datasets representative of the dynamic human body. Additionally, as health is inherently dynamic—subject to influences such as aging, lifestyle changes, and environmental factors—digital twins can be continuously updated to deliver an accurate and current overview of a patient's well-being. This ongoing refinement enables AI models that integrate digital twin technology to be much more adaptive and precise than ever before.

The integration of HDTs into everyday medical practices offers substantial benefits [76]. For example, they promise continuous patient monitoring that can predict potential infections and immune responses long before they escalate into emergencies. Moreover, they facilitate more efficient clinical trials by enabling simulations of new pharmaceuticals, which accelerates medication and vaccine development in safe, controlled virtual environments. On a patient level, these technologies allow for the crafting of personalized therapies where treatment plans can be tailored based on a patient's unique genetic makeup, thereby significantly reducing the risk of adverse side effects [76], [30].

Furthermore, the potential future applications of digital health twins extend well beyond current use cases. The ability to track and anticipate changes in a patient's health can enable early disease detection, particularly when biomarkers are employed to signal the onset of conditions like cancer. This early warning capability not only supports prompt preventive measures but also plays a critical role in minimizing healthcare disparities. For instance, HDTs coupled with telemedicine can ensure that high-quality care reaches patients regardless of their geographical location, making advancements in healthcare technology accessible to broader populations. Additionally, the quantification of subjective symptoms, such as pain and anxiety, into measurable data better equips healthcare providers to deliver accurate diagnoses and truly personalized treatment options [56].

In essence, digital twin technology represents a significant paradigm shift in our approach to healthcare. It is not merely enhancing existing medical protocols; it is establishing the groundwork for a future in which healthcare is more predictive, precise, and personalized. Ultimately, this evolution is expected to improve patient outcomes and facilitate a more efficient allocation of healthcare resources [48].

## 2.1 Applications of DTs in Healthcare

DTs can be implemented for many purposes, including the following in healthcare:

(1) **Proactive Disease Diagnosis:** Digital twins continuously collect patient data via IoT sensors, enabling the early identification of subtle biomarker changes that precede clinical symptoms. For instance, Allen et al. [7] demonstrated that machine learning models built on DT simulations can forecast disease progression—such as in stroke patients—by effectively mimicking patient physiology in near-real-time. Additionally, the architectural frameworks designed for healthcare digital twins provide robust integration of sensor data and advanced analytics to detect emerging clinical trends well before disease fully manifests [67]. This continuous monitoring facilitates a shift from reactive to proactive care.

(2) **Personalized Treatment:** By capturing individualized physiological and pathological details, digital twins allow for the simulation of various treatment scenarios tailored to a patient's unique profile. Gonsard et al. [32] exemplify this approach by demonstrating how DT models can be used to assess therapeutic interventions in chronic lung diseases, enabling treatment plans that are both personalized and adaptive. Furthermore, the life course digital twin model proposed by Milne-Ives et al. [63] underscores the potential of continuously refined simulations to optimize medication regimens and adjust therapies dynamically in response to patient-specific responses. This individualized strategy improves overall treatment efficacy and minimizes risks associated with a one-size-fits-all approach.

(3) **Preparatory Surgery:** The use of DTs allows the creation of virtual replicas of patient anatomy, providing surgeons with a rehearsal platform to practice intricate procedures before entering the operating room. Such simulation environments offer the opportunity to identify potential complications and refine surgical approaches in advance [67]. Although dedicated studies on these topics are still emerging, the ability to virtually simulate the patient's unique anatomical and physiological profile is increasingly recognized as a critical step in enhancing surgical safety and planning.

(4) **Vaccine Development and Usage:** Digital twins offer a simulated environment where vaccine candidates can be tested under various immunogenic conditions. Mariam et al. [58] discuss how integrating generative AI with DT technology can simulate diverse genetic and immunologic landscapes, enabling rapid evaluation of vaccine safety and effectiveness. Through these in silico trials, researchers can optimize dosages and fine-tune formulations, thereby accelerating the drug discovery process while reducing the risk of adverse effects during clinical trials.

(5) **Personalized Preventive Care:** By continuously analyzing fine-grained patient health data, digital twins empower clinicians to forecast potential health risks and implement targeted interventions before overt disease develops. Milne-Ives et al. [63] have shown that life course DT models can identify risk factors for multimorbidity, allowing for early, individualized preventive measures. In parallel, the structural designs for healthcare digital twins facilitate personalized recommendations, ranging from lifestyle adjustments to early clinical interventions, thereby reducing long-term healthcare burdens [67].

(6) **Optimizing Hospital Operations:** Beyond direct patient care, digital twins are instrumental in simulating and optimizing hospital operations. Virtual models of hospital systems—including patient flow, resource allocation, and staffing—enable administrators to test various operational scenarios and implement efficient solutions that reduce costs and improve crisis management. The DT framework for healthcare environments by Noeikham et al. [67] supports this multi-layered approach by integrating real-time data streams with simulation-based analytics, thus enabling decision-makers to address operational challenges before they manifest in practice. This approach is already influencing practices at healthcare institutions where companies like GE Healthcare and Siemens Healthineers are exploring DT-driven optimization strategies [15].

## 2.2   Characteristics of HDT

HDT is linked to its real-world counterpart through a unique identifier and continuously updated by real-time data collected from various medical and personal sources [87]. The following discussion outlines the features of an HDT:

(1) A primary attribute of the HDT is its unique identification and lifelong link. Each individual is associated with a singular digital identifier that functions as both an account login and as a persistent connection between the physical and virtual selves. This unique identification is critical for ensuring the continuity and integrity of the digital twin over time, and protocols have been developed to establish such persistent links within healthcare systems [53]. These identifiers also facilitate the secure correlation of personal data accumulated from diverse sources throughout an individual's life.

(2) Equally significant is the integration of genetic information into HDTs. While the creation of a digital twin at the moment of birth with inherited biological characteristics from ancestors is a conceptual framework, this process remains largely theoretical. Current practices focus on the aggregation of biometric and clinical data without formalized protocols for inheritance. Nevertheless, research indicates that the integration of multimodal diagnostic data—including genetic profiles—is feasible and can offer valuable insights into familial health traits [26].

(3) Once established, the HDT functions as a synchronous reflection of reality. It dynamically mirrors internal physiological states, including growth, aging, and responses to environmental factors such as medical treatments or injuries. This dynamic mirroring is essential for achieving near-real-time predictive analytics and early warning systems.

(4) Another distinguishing feature is data integration from a multitude of medical sources. With patient consent, data from various examinations, treatments, imaging studies, and wearable sensors are aggregated into the HDT. This comprehensive data collection creates a holistic view of the individual's health, enabling the integration of textual records, numerical measurements, and even imaging data. Sensor-enabled frameworks for digital twins in smart healthcare environments demonstrate such data assimilation and integration [4].

(5) Autonomous growth and predictive capabilities further empower the HDT. By programming attributes such as height, weight, and other growth parameters to update automatically based on established biological models, the HDT can continuously adapt based on incoming data from smart devices. Moreover, predictive analytics—integrated through hybrid cloud-edge architectures—enable the HDT to forecast potential health issues, supporting the delivery of proactive and personalized care [41].

(6) The HDT also serves as a tool for health assessment and personalized insights. Leveraging technologies such as Big Data analytics, AI, and expert systems, the digital twin can analyze its dataset to deliver tailored health evaluations, diagnoses, and therapeutic recommendations. This approach is central to the advancement of personalized medicine, as digital twins provide nuanced health assessments that cater to individual physiological variations [20].

(7) Interactive visualization and accessible interfaces are further characteristics that enhance the value of HDTs. Authorized users—whether patients, caregivers, or medical professionals—can log in to the HDT system to explore detailed 3D representations of the individual's health status, sometimes employing virtual or augmented reality technologies [35], [73]. Such immersive visualization offers intuitive diagnostic insights and aids in clinical decision-making [105].

(8) Finally, privacy, access control, and secure system infrastructure are paramount. Given the sensitivity of the integrated personal health data, HDT systems incorporate multilayered security measures, including biometric authentication, encryption, and secure communication protocols. Comprehensive surveys of digital twin architectures underscore the challenges and methodologies for ensuring robust data protection in these environments, emphasizing the importance of secure, trusted infrastructures in healthcare applications [101].

## 2.3 Comparison Between HDT and Conventional DT

This comparison can be understood by examining both their shared foundational characteristics and their crucial points of divergence:

### 2.3.1 *Similarities*.

- Both DTs and HDTs serve as virtual representations of physical entities.
- Fundamental to both approaches is the notion of a model-centric design wherein:
  - A computational model replicates the real world.
  - This enables simulations, diagnostics, and decision-support functions [88], [78].
- Both paradigms rely on bidirectional data flow:
  - This continuous exchange of information—from physical sensors to digital models and back to the physical system—ensures synchronization of the virtual representation with its real-world counterpart [88], [78].
- Underpinning these models is a common technological framework comprising:
  - Cloud computing infrastructures.
  - High-speed network communications.
  - Data acquisition systems.
  - Advanced analytics methodologies that facilitate data storage, processing, and real-time simulation [78].
- In both conventional DTs and HDTs, robust security mechanisms are critical for:
  - Safeguarding sensitive information.
  - Ensuring data integrity.
  - Maintaining system reliability [78].

### 2.3.2 *Differences*. Despite these similarities, HDTs introduce several additional layers of complexity due to their focus on replicating the intricacies associated with human beings.

- First, whereas DTs primarily model physical systems without subjective states, HDTs must incorporate cognitive and affective factors:
  - Capturing human emotions, mental states, and neurophysiological responses often necessitates the integration of specialized biosensors and brain–computer interface technologies—a requirement largely absent in conventional DT applications [74], [96].
- Second, human responses are inherently subjective and can vary widely across individuals:
  - This necessitates the integration of principles from psychology and behavioral science into HDT models [74], [96].
- Another key difference is the intrinsic complexity of humans as biological systems:
  - The development of an HDT requires interdisciplinary knowledge from biology, physiology, and psychology alongside conventional engineering principles.
  - HDTs must also model genetic inheritance and inter-individual genetic variations, requiring sophisticated data structures to represent hereditary information—a feature not needed in DT implementations [74].
- Further divergence is evident in the social and ethical considerations intrinsic to HDTs:
  - HDTs must address issues such as privacy, informed consent, and data ethics, as well as model interpersonal relationships and social dynamics—ethical dimensions generally peripheral to DTs focused on inanimate or simpler systems [27].
- Finally, HDTs must exhibit expanded environmental sensitivity:
  - They need to account for a wide range of influences—from physiological changes due to environmental exposures to psychological stressors, whereas conventional DTs are typically designed to monitor more narrowly defined physical parameters [74], [96].

## 3  UNIVERSAL FRAMEWORK OF THE HUMAN DIGITAL TWIN

The HDT's universal framework is composed of the following five interdependent components [21], [22], [62], [69], [99], [96], [80]:

### 3.1  Data Acquisition

In this stage, a continuous and diversified flow of information is collected from multiple sources. Medical records—such as electronic health records (EHR), diagnostic images, and biometric test results—are amalgamated with physiological measurements obtained from wearable devices (e.g., heart rate, blood pressure, and biomarker levels) to form a complete picture of a patient's health status [96], [80]. Additionally, the incorporation of social media activity extends the capture of emotional and psychological factors, reflecting an increasingly holistic approach to patient monitoring and personalized care [80]. This multifaceted data acquisition process is conceptually aligned with digital twin methodologies that demand the integration of heterogeneous data streams to construct accurate virtual replicas [80].

### 3.2  Digital Modeling and Virtualization

This component involves transforming the acquired data into a virtual twin (VT) that precisely replicates the physical twin (PT). Advanced digitization techniques convert body geometry, properties, and behavioral patterns into a digital model that is continuously updated with real-time information [96], [80]. The VT not only mirrors the patient's current state but is also made accessible through immersive interfaces—such as extended reality (XR) or holographic displays—thus facilitating intuitive and interactive engagement by healthcare professionals, patients, and caregivers [96]. The continuous conversion and live synchronization process underpin the fidelity of the HDT, demonstrating the robustness of digital twin implementations [96].

### 3.3  Communication and Computation

Efficient communication and computation form another critical pillar in the HDT ecosystem. A robust infrastructure is necessary to support the flow of information between the PT and VT, as well as between multiple digital twins. This involves establishing secure, high-capacity networks capable of handling bi-directional data exchanges in real time. Advanced processing techniques, including AI–based algorithms and edge computing, ensure prompt analysis and secure transmission of data, thereby maintaining an accurate digital representation of the physical entity [96], [80].

### 3.4  Data Management:

Data management is also paramount, given the vast volume and diversity of information collected from various sources. An effective data management system is needed to organize, filter, and store incoming data to maintain the VT's reliability and accuracy [96], [80]. Structured data management ensures that irrelevant or noisy inputs are minimized and that the data remain available for further analysis. This rigorous handling of data not only improves the quality of the digital model but also supports its evolution over time as new data feed into the system [96].

### 3.5  Data Analysis and Decision-Making

Finally, the HDT framework leverages advanced data analysis and decision-making tools to transform raw data into actionable insights. Sophisticated analytics—enabled by AI and ML—extract significant patterns from large clinical datasets, thereby providing healthcare providers with precise, data-driven recommendations [80]. This analytical capability is essential for supporting personalized treatment strategies, proactive diagnosis, and overall improved patient outcomes.

## 4 DESIGN REQUIREMENTS

The concept of HDTs for personalized healthcare (PH) captures many of the challenges and design requirements discussed in the literature on digital twin systems. Recent studies indicate that HDTs extend the traditional digital twin paradigm—originally applied in manufacturing, urban planning, or energy systems—to the healthcare domain by demanding high data fidelity, reliable and low-latency communication, advanced computational power, and robust data security protocols [24], [95], [64]. These scenarios introduce a strict set of design criteria and challenges for HDT, which are discussed below:

(1) **High-Quality Data:** This is the cornerstone of an accurate digital twin, and HDTs are no exception. In clinical applications, where real-time patient monitoring, individualized diagnoses, and personalized prescriptions are critical, the continuous acquisition of precise, large amounts of data (from both physiological sensors and environmental inputs) is essential. This need parallels findings in structural integrity management studies, where data quality assurance is crucial for system accuracy [54], and is further supported by many reviews on methodological challenges faced by DT systems in healthcare [60] and sensor-enabled frameworks [4]. Any degradation in data quality could lead to a mismatch between the patient's true state and its digital counterpart, jeopardizing clinical decisions.

(2) **Extreme Ultra-Reliable and Low-Latency Communication (xURLLC):** A significant challenge is the necessity for an extreme ultra-reliable and low-latency communication (xURLLC), meaning speeds of at least 100 Gbps, almost perfect reliability (99.99999%), and delays of 1 millisecond or less. Research employing deep reinforcement learning has explored methods to achieve scheduling with round-trip times (RTTs) as low as 1 millisecond [106]. For time-critical procedures such as remote surgery, even marginal delays can have serious consequences. Studies on IoT-based healthcare systems highlight that current 5G technologies do not fully meet the rigorous standards required for such applications [42]. These studies collectively advocate for the development of next-generation communication infrastructures designed specifically for HDTs.

(3) **Ultra-Low Round-Trip Time (RTT):** In specific healthcare scenarios (e.g., surgical simulations or interactive therapies), the demand for ultra-low RTT reinforces the idea that HDT computing resources should be located as near as possible to the end user. This architectural recommendation aligns with research on ultra-reliable low-latency systems, where integrating edge computing is suggested to surpass the limitations of centralized data processing [106].

(4) **Data Privacy, Security, and Integrity:** HDTs involve not only clinical and personal health information but also real-time status data from wearable sensors, necessitating strict adherence to data protection standards. Blockchain-based approaches have been proposed to secure patient digital twins against unauthorized access and tampering [8]. Similar concerns are noted in discussions about the methodological challenges for digital twins in healthcare [60] and in AI-driven frameworks that emphasize data security [1], [19].

(5) **Data Storage and Computing Power:** The large volume of data generated by HDTs—often several gigabytes per day—requires robust cloud and distributed storage solutions along with advanced computing infrastructures capable of real-time analysis and model updates. These infrastructures ensure that the HDT remains a faithful and up-to-date representation of the patient while enabling complex decision-making algorithms.

(6) **AI-Driven Analytics:** AI-driven methods analyze incoming data continuously to predict healthcare outcomes and provide actionable insights. However many AI systems operate as "black boxes," hindering clinician trust and complicating regulatory approval. This call for interpretability reinforces the need for ongoing research into AI methodologies that adapt in real-time to changing patient conditions.

Realizing the potential of HDTs in PH hinges on an end-to-end networking architecture that mirrors the operational workflow of digital twins. A well-structured architecture typically includes five layers: data acquisition, communication, data management (where pre-processing, storage, and sharing occur), computational processing, and data analysis/decision-making—with feedback to the physical twin. This layered approach is conceptually sound and has been successfully implemented in IoT-based healthcare frameworks [42], [55] and sensor-enabled digital twin systems [4]. This architecture ensures seamless data flow, robust real-time analysis, and continuous synchronization between the patient and their digital counterpart, addressing the multifaceted requirements of modern precision medicine. Further details regarding each layer will be presented in the subsequent sections.

## 5 DATA ACQUISITION LAYER

The data acquisition layer leverages advanced sensing technologies to collect a diverse range of physiological and psychological indicators from mobile individuals. In this layer, data is obtained by integrating signals from multiple heterogeneous sources, including Internet of Medical Things (IoMT)-enabled sensors, soft sensors derived from social media, and Electronic Health Records (EHRs) [75], [52].

### 5.1 IoMT-Enabled Sensors

Wearable biomedical devices are designed for continuous, unobtrusive monitoring and are categorized by their placement on the body:

*5.1.1 Head-Mounted Devices.* Devices like smart glasses can track habits and real-time ECG signals. Smart contact lenses monitor biochemical parameters, such as glucose levels [90], [52].

*5.1.2 Torso-Worn Sensors.* Smart garments, skin patches, and sensor-integrated temporary tattoos record vital signs including pulse rate, EEG signals, and subtle sweat biomarkers.

*5.1.3 Limb-Worn Accessories.* Accessories such as smart armbands and smart apparel (pants, shoes) capture data on physical activity and gait dynamics through sensors like photoplethysmography (PPG) and accelerometry. These devices are crucial for managing conditions like diabetes [18], [14].

*5.1.4 Implantable Sensors.* Recent advances in nanotechnology have led to the development of implantable sensors that provide precise measurements of internal conditions (e.g., intracranial pressure, temperature), enhancing early disease detection and continuous health evaluation [75].

### 5.2 Soft Sensors from Social Media

The data acquisition layer also utilizes "soft sensors" from platforms like Instagram, X, and Facebook. These platforms provide real-time insights into psychological and emotional states through sentiment analysis of user-generated content [111].

### 5.3   Electronic Health Records (EHRs)

EHRs contribute historical and diagnostic data—such as medical images, diagnoses, allergy information, and treatment records—establishing a robust clinical baseline for constructing human digital twins. This integration enhances personalization in diagnostic and therapeutic strategies and supports improved clinical decision-making [98].

**Challenges**

Despite advances, significant challenges persist:

- **Power Limitations:** Many wearable and implantable devices operate under strict battery constraints, hindering continuous monitoring capabilities [49].
- **Data Integration:** The heterogeneous data collected at varying temporal resolutions complicates the creation of a coherent HDT model, necessitating advanced data fusion techniques [47].
- **Data Management:** The massive volume of sensitive health data requires robust systems with rigorous security and privacy protections, resulting in the need for innovations in data storage and ethical frameworks [16, 86].

## 6   COMMUNICATION LAYER

The communication layer in HDT systems is architected into two distinct tiers. The first is dedicated exclusively to on-body data exchange, while the second enables beyond-body networking to connect wearable and implantable sensors to remote virtual servers:

(1) **On-Body Communication:** On-body communication relies on Wireless Body Area Networks (WBANs) to collect physiological and psychological data from various low-power sensors positioned on or near the human body. In these networks, sensors—often integrated into wearable devices such as smartwatches or dedicated monitoring patches—transmit data either directly to a central hub (typically a smartphone) in a star topology or initially to a local processor in a hierarchical topology. The use of a star topology minimizes communication overhead by allowing each sensor to interface independently with the gateway, whereas the hierarchical structure facilitates preliminary data processing that reduces network load and helps conserve energy [77], [68]. Because sensor nodes are constrained by battery and energy-harvesting capabilities, the choice of communication protocols is critical.

   - Bluetooth Low Energy (BLE) is often employed due to its low power consumption and adequate short-range performance; however, its lack of native multicast support can limit certain network operations [77], [93].
   - ZigBee, which is based on the IEEE 802.15.4 standard, offers a flexible network configuration that can support a larger number of devices with built-in security measures. Nonetheless, ZigBee implementations may suffer from interference and are vulnerable to radio jamming, highlighting the need for further enhancements in robustness [102], [51].
   - In addition to these conventional electromagnetic methods, Molecular Communication (MC) has emerged as an alternative inspired by biological processes. MC holds promise for in-body scenarios where electromagnetic propagation is inefficient, although challenges remain in interfacing internal biological environments with external systems and developing reliable multi-channel (MIMO) MC approaches [108], [34].
   - Complementary solutions such as Narrowband IoT (NB-IoT) and IPv6 over Low-Power Wireless Personal Area Networks (6LoWPAN) are also being explored to expand the range of energy-efficient communication options in this tier [94], [70].

(2) **Beyond-Body Communication:** Beyond-body communication is responsible for linking local gateways and central hubs to remote servers (e.g., cloud data centers) that host the VT of the patient. This tier supports the bi-directional and real-time exchange of large volumes of multimodal data—including text, images, video, and haptic feedback—which is crucial for latency-sensitive applications such as remote surgery simulation. Two notable technologies in this space are:

- Cutting-edge approaches such as the Tactile Internet (TI) have been engineered to deliver ultra-reliable low-latency communication, ensuring that even haptic data can be transmitted almost instantaneously [108], [34].

- Additionally, semantic communication is gaining traction as an emerging paradigm that reduces bandwidth requirements by transmitting only the semantically meaningful components of data. By leveraging advanced machine learning and knowledge representation techniques, semantic communication not only reduces the overall data volume but also improves energy efficiency and enhances security through contextual data interpretation [108], [34].

**Challenges**

Still, several challenges persist across both tiers of the communication layer:

- **Interference:** The dense deployment of heterogeneous wireless devices operating in overlapping frequency bands can result in co-channel interference that compromises signal integrity. To address these issues, various interference mitigation strategies based on sophisticated coding techniques and media access control protocols have been proposed [107].

- **Resource Optimization:** This represents another major challenge; ensuring minimal transmission delay and maintaining energy efficiency require adaptive medium access control protocols and routing algorithms that can dynamically balance network resources, particularly in time-sensitive healthcare applications [71], [5].

- **Security and Privacy:** Finally, the security and privacy of the data transmitted via these networks are of critical importance given the sensitive nature of healthcare information. Robust security frameworks that incorporate encryption, authentication, and secure key management are essential to safeguard patient data both within the WBAN and during its transmission to external servers [6], [39].

## 7  COMPUTATION LAYER

The evolution of digital twin technology has underscored the need for computational platforms that exceed the capabilities of typical mobile or IoT devices. In particular, applications such as real-time physical twin rendering and data-driven interactions demand extremely low latency and high reliability that traditional cloud infrastructures alone cannot guarantee. Two advanced paradigms—Multi-Access Edge Computing (MEC) and Edge-Cloud Collaboration—have emerged to address these requirements.

(1) **Multi-Access Edge Computing (MEC):** MEC extends cloud-like resources to the network's periphery by deploying computational capabilities at locations such as base stations and access points [29], [91]. This proximity offers multiple benefits for HDT systems:

- **Supporting Mobility:** This architectural design allows processing tasks to be executed in proximity to the PT, which is essential for maintaining low latency amidst user mobility. Studies have demonstrated that by handling computation near the data source, MEC supports mobile applications that require continuously adapting VT—a necessity for real-time healthcare monitoring and personalized coaching, where delay sensitivity is critical [91], [103].

- **Accelerating Response Times:** The minimization of the transmission distance between end-users and processing nodes substantially accelerates response times. Rapid feedback loops are indispensable in applications such as cardiac monitoring and real-time athletic training, and the reduced end-to-end delay offered by MEC has been corroborated [29], [33].
- **Ensuring High Quality-of-Service (QoS):** MEC meets the rigorous QoS demands of HDT, providing ultra-low latency and high reliability. Its ability to guarantee ultra-low latency and high reliability ensures the quality of service (QoS) demanded by immersive augmented reality (AR) scenarios and other interactive applications [29], [91].

Despite these benefits, MEC faces challenges, including high energy consumption and limited computational capacities at edge servers, as well as potential data privacy risks. To address these issues, ongoing research focuses on improved task scheduling, dynamic resource allocation, and multi-user offloading strategies, with energy efficiency being a particularly prominent concern [9], [25].

(2) **Edge-Cloud Collaboration:** While MEC is well-suited for low-latency tasks, its inherent resource constraints can limit its effectiveness for complex computations—tasks such as detailed biological simulations often require the vast processing power available only in centralized cloud environments. To bridge this gap, edge-cloud collaboration integrates the rapid response capabilities of MEC with the expansive resources of cloud computing [109]. This process will be handled in the following manner:

- **Optimized Computation and Latency Management:** In such a hybrid framework, time-critical tasks are processed at the edge, where minimal delay is paramount, whereas computationally intensive processes are offloaded to the cloud. This strategic partitioning minimizes overall latency while making optimal use of heterogeneous resources [109].
- **Improved Data Distribution and Enhanced Privacy:** By distributing data processing over both edge and cloud infrastructures, the approach helps reduce potential breach points for sensitive information. Techniques such as Federated Learning (FL) enable local training of AI models without the need to share raw data externally; the locally derived insights can then be aggregated at the cloud level, thus enhancing privacy protection while capitalizing on collective computational power [66].

Despite these clear advantages, the edge-cloud approach also introduces challenges such as effective task partitioning, distributed resource management, and robust security and access control measures to ensure system integrity and data safety [109], [66].

## 8   DATA MANAGEMENT LAYER

This layer addresses challenges arising from continuously generated heterogeneous data streams. In such systems, data originate from pervasive sensors (e.g., wearable and implantable devices that capture physiological metrics), social networks that provide behavioral and emotional insights, Electronic Health Records (EHRs) containing extensive historical medical information, and VTs that generate simulation outputs and feedback. This multiplicity of sources inherently creates data characterized by a high degree of variety, volume, dimensionality, noise, and sensitivity. The heterogeneous nature of these inputs necessitates robust processing and management techniques, as underscored in contemporary research on sensor-based healthcare systems and big data analytics [97], [11].

To effectively harness such diverse and high-velocity data, the data management layer is designed to extend beyond mere storage functions by incorporating a three-step process: data pre-processing, data storage, and data security and privacy.

(1) **Data Pre-processing:** This stage is crucial for enhancing data quality prior to analysis. Before data is used, it must be cleaned and prepared. This involves several key activities:

- First, data cleaning is performed to remove noise, outliers, and other anomalies, with imputation techniques—such as k-nearest neighbors (KNN) or deep learning–based methods—used to manage missing values [3].

- Next, data reduction is achieved through feature selection and feature extraction. Simultaneous techniques that select only the most significant data attributes, as well as methods that transform raw data into more compact representations, have proven effective in improving the performance of machine learning models in healthcare applications [92], [57].

- In addition, data fusion integrates outputs from multiple sources either at the raw-data, feature, or decision levels, thereby forming a coherent and enriched dataset that leverages complementary information across heterogeneous sources [92].

(2) **Data Storage:** Following pre-processing, the data storage phase must provide robust and scalable solutions to support real-time analytics. Distributed architectures—such as the Hadoop Distributed File System (HDFS) for managing vast datasets, NoSQL databases like HBase for supporting rapid read/write operations, and object storage systems such as OpenStack Swift for unstructured data—are widely employed to meet these demands [97], [45].

(3) **Data Security and Privacy:** Finally, given that the data often encompass highly sensitive personal health information, the data management layer must incorporate comprehensive security and privacy measures.

- Cybersecurity strategies, including intrusion detection systems, are implemented to guard against threats such as denial-of-service attacks, data injections, and eavesdropping [97].

- In parallel, privacy-preserving techniques are deployed; these include a range of cryptographic methods (symmetric, asymmetric, homomorphic, and post-quantum encryption), anonymization protocols to remove personal identifiers, as well as federated learning and differential privacy, which enable collaborative model training without compromising raw data security [11].

- Moreover, distributed ledger technologies such as blockchain are being explored to create immutable and transparent logs of data transactions, thereby enhancing access control and ensuring data integrity in environments where trust and auditability are paramount [97].

## 9 DATA ANALYSIS AND DECISION MAKING LAYER

The data analysis and decision-making layer functions as the "brain" that transforms vast amounts of raw physiological and psychological data into actionable insights to support PH applications. Essentially, AI serves as the engine for this transformation by detecting complex patterns, forecasting future health states, supporting clinical decisions, and continuously evolving the VT to mirror the PT [36].

The primary functions in this layer include:

- **Pattern Recognition:** Detecting trends, anomalies, and correlations in physiological and psychological data.
- **Prediction:** Estimating future health conditions and identifying potential risks.

- **Decision Support:** Offering guidance on diagnosis, treatment options, and personalized interventions.
- **Model Evolution:** Ensuring that the VT stays updated and accurately mirrors the PT.

Three main types of AI are used to accomplish these goals:

### 9.1 Supervised Learning

Among the AI techniques adopted, supervised learning is frequently employed because it trains models with data where the outcomes are known. Key algorithms such as KNN and Support Vector Machines (SVM) are widely applied for tasks like disease detection [104]. Furthermore, deep learning techniques—including convolutional neural networks (CNNs)—automate feature extraction from complex data, as demonstrated in studies for COVID-19 diagnosis and electrocardiogram (ECG) analysis [82], [81], [2].

### 9.2 Unsupervised Learning

In contrast, unsupervised learning methods are utilized when there are no predefined outcomes, enabling the discovery of hidden structures within data. Techniques like K-Means Clustering have been used for grouping patients with similar health characteristics, and generative adversarial networks (GANs) promote data augmentation and anonymization. These methods have proved particularly effective for patient stratification and anomaly detection in physiological data, as highlighted in recent investigations [38].

### 9.3 Reinforcement Learning

Furthermore, reinforcement learning (RL)—where an agent learns to make sequential decisions by maximizing a long-term reward—has begun to show promise in the healthcare domain. For instance, RL strategies can be applied for tasks like personalized prescription management and surgical planning. Although RL is still an emerging area in clinical settings, its incorporation is driven by the need for adaptive and continuously improving decision support in complex healthcare environments. This integration ultimately ensures that the VT remains an accurate, real-time representation of the PT while accommodating individual patient dynamics.

### Challenges

By transforming cleansed data into insights via AI algorithms, this layer not only paves the way for personalized recommendations but also faces challenges that need addressing. Explainability remains a significant issue—for many models are "black boxes" whose internal decision-making processes are not transparent—and training robust models is often hindered by data scarcity, heavy computational demands, and generalization issues. These challenges underline the need for further research in explainable AI, data augmentation, and model optimization to build trust and ensure effectiveness in clinical applications [36].

## 10 OPEN ISSUES

Despite the high potential of DTs in healthcare, their development and implementation face several significant hurdles:

### 10.1 Ethical and Policy Concerns

- **Data Security and Privacy:** Implementing DTs demands gathering extensive sensitive patient data, which raises ethical questions about confidentiality and protection. There is a risk that detailed health information could be misused—such as by insurance companies adjusting premiums based on indicators like physical activity or diet patterns—and even expose systems to cyber-attacks. Additionally, new legal standards, like those in the EU's GDPR, mandate strict data handling procedures and grant patients rights such as the ability to withdraw consent or request deletion of their records [100].

- **Bias in Data:** Successful DT applications depend on balanced datasets. Many current healthcare records show demographic imbalances, leaning towards overrepresented groups such as white males. This skew can lead to recommendations that fail to serve underrepresented patients effectively.

- **Genetic Profiling Risks:** DTs can highlight genetic traits associated with better health outcomes. While beneficial, this capability could be misused to favor certain genetic profiles over others, fueling discriminatory practices and risky policies based on the idea of a "superior" gene pool.

- **Equity in Access:** Uncertain business models could limit DT-based treatments to those who can afford them, leaving behind lower-income patients. Without broad coverage for insurance or public integration, these innovations could exacerbate existing healthcare disparities [100].

### 10.2 Technical Limitations

- **Emerging Technical Challenges:** Many current barriers in DT technology align with ongoing research areas. Overcoming these will be essential to enhance both reliability and integration into clinical workflows.

### 10.3 Social and Professional Barriers

- **Skepticism Towards Algorithmic Decisions:** Clinicians often remain cautious about relying on automated predictions [79]. Concerns persist about the opaque nature of some algorithm-based decisions, which may lack clear explanations and thereby risk misdiagnoses or improper treatments [10].

- **Fear of Professional Displacement:** Although DTs can process vast amounts of data quickly—sometimes surpassing human capabilities—this advantage has sparked concerns about replacing clinicians. In practice, current DT solutions are more likely to serve as supportive tools that integrate seamlessly with existing medical workflows rather than replace human judgment.

## 11 FUTURE DIRECTIONS

HDTs hold enormous promise for PH, but several critical barriers must be overcome. These challenges span technical, ethical, and operational dimensions, and addressing them will require innovative, cross-disciplinary research. Here are the key directions that research should pursue:

### 11.1 Overcoming Data Scarcity

- **Challenge:** HDTs need vast, diverse, and high-quality datasets, yet data is often fragmented, unstandardized, and limited for rare diseases or underrepresented populations.

- **Research Focus:**
  - **Federated Data Sharing:** Develop secure frameworks (using technologies like blockchain and differential privacy) to allow institutions to share data without exposing sensitive details.
  - **Standardization:** Establish common data protocols and ontologies to seamlessly merge diverse datasets.
  - **Synthetic Data Generation:** Utilize AI methods such as GANs to create realistic datasets that complement limited real-world data, ensuring rigorous validation to avoid bias.

    – **Data Minimization:** Only collect the data necessary for effective modeling.

## 11.2 Supporting Mobility in HDTs

- **Challenge:** As people move, maintaining a seamless and low-latency connection between the PT and its virtual counterpart is difficult.
- **Research Focus:**
  - **Adaptive Migration:** Create algorithms to dynamically reposition virtual twins based on patient location and network conditions.
  - **Handover Protocols:** Design technologies to ensure uninterrupted data flow during movement.
  - **Edge-Cloud Integration:** Leverage the low latency of edge computing for real-time processing while using cloud resources for heavier computations.
  - **Energy Efficiency:** Optimize protocols to reduce battery consumption on mobile devices.

## 11.3 Advancing Federated HDTs

- **Challenge:** Single-server hosting is often insufficient for the complex operations of HDTs, necessitating distributed solutions that introduce issues of consistency and resource management.
- **Research Focus:**
  - **Distribution Strategies:** Develop methods to replicate and distribute virtual twins across multiple nodes while considering data locality and network topology.
  - **Efficient Synchronization:** Implement robust techniques to keep data consistent across decentralized systems.
  - **Collaborative Mechanisms:** Explore approaches such as model averaging and transfer learning to boost performance.
  - **Dynamic Resource Orchestration:** Design algorithms that allocate computing resources intelligently across heterogeneous nodes.
  - **Enhanced Security:** Intensify research into privacy and security in distributed settings.

## 11.4 Achieving Interoperable Subsystem Management

- **Challenge:** Modeling the interrelated subsystems of the human body (e.g., circulatory, respiratory) into a single, cohesive HDT is inherently complex.
- **Research Focus:**
  - **Semantic Interoperability:** Create frameworks that use standard ontologies for effective data exchange between subsystem models.
  - **Model Integration:** Investigate methods to combine models from various physiological systems into a unified HDT.
  - **Scalable Architectures:** Develop management systems that can coordinate multiple subsystems efficiently.

## 11.5 Designing User-Centric Interfaces

- **Challenge:** Interfaces must be intuitive, secure, and capable of handling complex HDT data for various stakeholders.
- **Research Focus:**

  – **Human-Centered Design:** Prioritize ease of use and tailored interactions for patients and professionals.
  – **Multimodal Interaction:** Support varied input methods (voice, gesture, touch) for better engagement.
  – **Robust Security:** Ensure that all interfaces incorporate strong data protection and authentication measures.
  – **Scalability and Standardization:** Develop interfaces that perform well at scale and adhere to universal standards for easy integration.

## 11.6  Integrating Intelligent Blockchain Solutions

- **Challenge:** Traditional blockchain systems can be too slow and resource-intensive for real-time HDT applications.
- **Research Focus:**
  – **Lightweight Consensus:** Create fast and secure consensus mechanisms.
  – **Scalability Improvements:** Design blockchain architectures capable of managing high data throughput.
  – **Edge Integration:** Combine blockchain with edge computing to reduce latency.
  – **Smart Contract Automation:** Use smart contracts for automated management of data access and system updates.

## 11.7  Enhancing Explainable AI (XAI)

- **Challenge:** Many AI systems in HDTs operate as "black boxes," reducing trust and transparency.
- **Research Focus:**
  – **Explainable Techniques:** Integrate XAI methods to clarify how AI-driven decisions are made.
  – **Human-in-the-Loop:** Develop systems that incorporate user feedback to adjust and improve predictions.
  – **Auditability:** Ensure AI processes are fully auditable to build accountability.

## 11.8  Developing Generalized AI

- **Challenge:** Current approaches tend to be data-hungry and overly specific, limiting applicability across different scenarios.
- **Research Focus:**
  – **Transfer and Meta-Learning:** Investigate methods to develop models that generalize better across tasks.
  – **Unsupervised Methods:** Explore training techniques that reduce reliance on large, labeled datasets.

## 11.9  Pursuing Green HDTs

- **Challenge:** The substantial computational requirements of HDTs can lead to high energy consumption and environmental impact.
- **Research Focus:**
  – **Energy-Efficient Systems:** Design architectures that reduce power usage.
  – **Green Technologies:** Embrace renewable energy and low-consumption hardware.
  – **Optimization Algorithms:** Develop methods to optimize resource utilization without sacrificing performance.

### 11.10  Integrating HDTs with the Metaverse

- **Challenge:** Offering immersive healthcare experiences in the metaverse comes with concerns about security, interoperability, and real-time data synchronization.
- **Research Focus:**
  - **Secure Platforms:** Ensure robust protection of HDT data with advanced encryption and authentication.
  - **Interoperability Standards:** Standardize data formats and protocols to facilitate interactions between different metaverse environments.
  - **Real-Time Synchronization:** Enhance network protocols to maintain an accurate and responsive digital twin experience.

## 12  CONCLUSION

This paper presented a detailed technical overview and networking architecture for Human Digital Twins (HDTs) in personalized healthcare (PH). I articulated the motivations for HDT adoption, highlighted its principal benefits over traditional digital twin (DT) models, and examined supporting technologies across several domains, including data collection methods, communication techniques, computing paradigms, data management processes, data analysis, and decision-making methods. Despite the solid technological underpinnings, several critical challenges remain. In particular, effective modeling of the complexities inherent in human physiology, integration of heterogeneous and dynamic data sources, and the fusion of multifaceted datasets necessitate innovative algorithmic approaches and scalable computing infrastructures. In parallel, addressing ethical, cultural, and patient safety concerns is essential to attain widespread acceptance. The rapid progression in sensor technologies, computational power, and digital twin methodologies inspires confidence that, with sustained interdisciplinary research, HDTs will transform personalized diagnosis, therapy, and patient management.